\documentclass[10pt,conference]{IEEEtran}

\usepackage{cite}

\ifCLASSINFOpdf
   \usepackage[pdftex]{graphicx}
   \graphicspath{{./pics/}}
   \DeclareGraphicsExtensions{.pdf,.jpeg,.png}
\else
   \usepackage[dvips]{graphicx}
   \graphicspath{{./pics/}}
   \DeclareGraphicsExtensions{.eps}
\fi
\usepackage[cmex10]{amsmath}
\usepackage{amsthm}

\usepackage{algorithmic}

\usepackage{array}

\ifCLASSOPTIONcompsoc
  \usepackage[caption=false,font=normalsize,labelfont=sf,textfont=sf]{subfig}
\else
  \usepackage[caption=false,font=footnotesize]{subfig}
\fi
\usepackage{balance, multirow, pstool, url}

\begin{document}
%
\thispagestyle{empty}
\onecolumn
\linespread{1.2}\selectfont{}
{\noindent\Huge IEEE Copyright Notice}\\[1pt]

{\noindent\large Copyright (c) 2018 IEEE

\noindent Personal use of this material is permitted. Permission from IEEE must be obtained for all other uses, in any current or future media, including reprinting/republishing this material for advertising or promotional purposes, creating new collective works, for resale or redistribution to servers or lists, or reuse of any copyrighted component of this work in other works.}\\[1em]

{\noindent\Large Accepted to be published in: 2018 31st SIBGRAPI Conference on Graphics, Patterns and Images (SIBGRAPI'18), October 29 -- November 1, 2018.}\\[1in]

{\noindent\large Cite as:}\\[1pt]

{\setlength{\fboxrule}{1pt}
 \fbox{\parbox{0.7\textwidth}{L. A. Duarte, O. A. B. Penatti and J. Almeida, ``Bag of Attributes for Video Event Retrieval,'' in \emph{2018 31st SIBGRAPI Conference on Graphics, Patterns and Images (SIBGRAPI)}, Foz do Igua\c{c}u, Brazil, 2018, pp. 447-454, doi: 10.1109/SIBGRAPI.2018.00064}}}\\[1in]
 
{\noindent\large BibTeX:}\\[1pt]

{\setlength{\fboxrule}{1pt}
 \fbox{\parbox{0.85\textwidth}{
 @InProceedings\{SIBGRAPI\_2018\_Duarte,
 
 \begin{tabular}{lll}
  & author    & = \{L. A. \{Duarte\} and O. A. B. \{Penatti\} and J. \{Almeida\}\},\\
			   
  & title     & = \{Bag of Attributes for Video Event Retrieval\},\\
			   
  & pages     & = \{447-454\},\\
  
  & booktitle & = \{2018 31st {SIBGRAPI} Conference on Graphics, Patterns and Images ({SIBGRAPI})\},\\
  
  & address   & = \{Foz do Igua\c{c}u, Brazil\},\\
  
  & month     & = \{October 29 -- November 1\},\\
  
  & year      & = \{2018\},\\
  
  & publisher & = \{\{IEEE\}\},\\
  
  & doi       & = \{10.1109/SIBGRAPI.2018.00064\},\\
  \end{tabular}
  
\}
 }}}

\twocolumn
\linespread{1}\selectfont{}
\clearpage

%
\title{Bag of Attributes for Video Event Retrieval}

\newif\iffinal
\finaltrue
\newcommand{\jemsid}{59}
\newcommand{\rv}[1]{#1}


\iffinal

\author{
	\IEEEauthorblockN{Leonardo A. Duarte$^1$, Ot\'{a}vio A. B. Penatti$^2$, and Jurandy Almeida$^1$}\\
	\IEEEauthorblockA{%
	  $^1$Instituto de Ci\^{e}ncia e Tecnologia\\
	  Universidade Federal de S\~{a}o Paulo -- UNIFESP\\
	  12247-014, S\~{a}o Jos\'{e} dos Campos, SP -- Brazil\\
	  Email: \small\texttt{\{leonardo.assuane, jurandy.almeida\}@unifesp.br}\\[1ex]
	  $^2$Advanced Technologies\\
	  SAMSUNG Research Institute\\
	  13097-160, Campinas, SP -- Brazil\\
	  Email: \small\texttt{o.penatti@samsung.com}}
}


%

\else
  \author{SIBGRAPI paper ID: \jemsid \\ }
\fi

\maketitle

\begin{abstract}
In this paper, we present the Bag-of-Attributes (BoA) model for video representation aiming at video event retrieval.
The BoA model is based on a semantic feature space for representing videos, resulting in high-level video feature vectors.
For creating a semantic space, i.e., the attribute space, we can train a classifier using a labeled image dataset, obtaining a classification model that can be understood as a high-level codebook. 
This model is used to map low-level frame vectors into high-level vectors (e.g., classifier probability scores). 
Then, we apply pooling operations \rv{to} the frame vectors to create the final bag of attributes for the video. 
In the BoA representation, each dimension corresponds to one category (or attribute) of the semantic space. 
Other interesting properties are: compactness, flexibility regarding the classifier, and ability to encode multiple semantic concepts in a single video representation.
Our experiments considered the semantic space created by state-of-the-art convolutional neural networks pre-trained on 1000 object categories of ImageNet. 
Such deep neural networks were used to classify each video frame and then different coding strategies were used to encode the probability distribution from the softmax layer into a frame vector. 
Next, different pooling strategies were used to combine frame vectors in the BoA representation for a video. 
Results using BoA were comparable or superior to the baselines in the task of video event retrieval using the EVVE dataset, with the advantage of providing a much more compact representation.
\end{abstract}


\IEEEpeerreviewmaketitle

\section{Introduction}
\label{sec:intro}
The retrieval of videos from specific events, e.g., the wedding of Prince William and Kate Middleton or the riots of 2012 in Barcelona, is a challenging application, as the goal is to retrieve other videos from that event 
in a database containing lots of different events. 
This task is even more challenging if we are considering only visual content, i.e., no textual annotations. 
Different events can occur at the same locations but in different dates, making videos of such events very similar visually. 
Other challenge is that there can be a large variation in visual aspects, even in the same event. 
For instance, for the wedding of Prince William and Kate Middleton, there can be videos with close-ups in the people and videos of the location (church, city buildings, etc).

Traditional video descriptors are usually based on low-level features, like textures and local patches~\cite{ICIP_2011_Almeida, CVPR_2011_Wang, ECCV_2008_Willems}, which rarely represent semantic properties. 
Some more recent approaches aim at including semantics in video representations~\cite{CIKM_2013_Dalton, IJCV_2014_Patterson, CVPR_2014_Wu}.
Action Bank~\cite{CVPR_2012_Sadanand} is a method to represent videos according to a bank of action detectors. 
Bag of Scenes~\cite{ICMR_2012_Penatti} considers a dictionary of scenes instead of a dictionary of local patches, and uses it for video geocoding, as the scenes can be representative of places. 
Works related to semantic signatures~\cite{WACV_2016_Agharwal, ICMR_2013_Habibian, MM_2013_Mazloom, TMM_2012_Merler} use concept detectors to obtain high-level video representations.
Many works are based on obtaining frame-level feature vectors and then aggregating them as a video vector~\cite{AAAI_2013_Jiang, CVPR_2013_Ma, CVPR_2015_Ng, MM_2015_Ye, CVPR_2015_Xu, MM_2015_Wu, BMVC_2015_Zha}.

Identifying and representing semantics of a video content is one of the most important aspects for video analysis, classification, indexing, and retrieval.
If we could have a representation that can encode the multiple elements that appear in a given event in a single feature vector, we could better describe such event and discriminate it from others. 
Such a representation can be achieved by considering a classifier of high-level concepts in the video. 
Such concepts could be objects, scenes, locations, and so on.

To achieve such high-level representation for video event retrieval, we present the Bag-of-Attributes (BoA) model.
The BoA model is based on a semantic feature space for representing video content, i.e., the attribute space, which can be understood as a high-level visual codebook.
This space can be created by training a classifier using a labeled image dataset.
Video contents can then be described by applying the learned classifier.
The video vector contains the responses of the classifier, in which we have the activations of the semantic concepts that appear in the video.
Such representation is a high-level feature vector for the video.

We validated the BoA model for video event retrieval using the EVVE dataset~\cite{CVPR_2013_Revaud}. 
For obtaining the semantic feature space, we used state-of-the-art convolutional neural networks (CNNs) pre-trained on 1000 object categories of ImageNet.
These deep neural networks were used to classify each video frame and then different coding strategies were used to encode the probability distribution from the softmax layer as a high-level frame vector. 
Next, different pooling strategies were applied over the frame vectors, creating the final video vector (i.e., the bag of attributes). 
Results point that the BoA model can provide comparable or superior effectiveness to existing baselines for video event retrieval, with the advantage of BoA generating more compact feature vectors.

The remainder of this paper is organized as follows. 
Section~\ref{sec:related} introduces some basic concepts and describes related work. 
Section~\ref{sec:boamodel} presents the BoA model and shows how to apply it for representing video data. 
Section~\ref{sec:results} reports the results of our experiments and compares our technique with other methods. 
Finally, we offer our conclusions and directions for future work in Section~\ref{sec:conclusions}.

\section{Background and Related Work}
\label{sec:related}
The BoA model is a high-level representation for video event retrieval. 
Thus, it is related to the areas of visual dictionaries, video representation, and to the task of video event retrieval. 
In the following sections, we give some background and present related work in each area separately.

\subsection{Visual Dictionaries}
\label{sec:visualdictionaries}
Visual dictionaries have been the state-of-the-art representation for many years in visual recognition. 
Bag of Visual Words, which are representations based on visual dictionaries, have the ability of 
preserving the discriminating power of local descriptions while pooling those features into a single feature vector~\cite{CVPR_2010_Boureau}.

To obtain a Bag-of-Visual-Words (BoVW) representation, the visual dictionary or codebook needs to be created, so the visual content of interest can be represented according to the visual dictionary.
Visual dictionaries are created by quantizing a feature space, usually using unsupervised learning approaches and based on a feature space of local features.
Such local features are extracted from a training set of images or videos.
For extracting local features, 
one can use interest point detectors or employ dense sampling in a regular grid to obtain local patterns to be described.
Then, each local pattern is described by local features, like SIFT and STIP.
Those local features are clustered or randomly sampled in order to obtain the visual words of the dictionary. 
The clustering of the feature space is based solely on the visual appearance and, hence, the visual words themselves carry no semantics~\cite{NIPS_2010_Li}.

Thereafter, the created dictionary is used to represent the visual content of interest. 
This is performed by a step usually called \textit{coding}, in which the local features of the content of interest are encoded according to the dictionary.
The coding step can employ \textit{hard} or \textit{soft} assignment, for instance. 
After encoding local features according to the dictionary, a \textit{pooling} step is applied to summarize the assignment values and generate the final feature vector. 
Popular pooling strategies are \textit{average} and \textit{max} pooling. 

As explained above, visual dictionaries are commonly based on unsupervised learning, therefore, having no explicit semantics.
The BoA model aims to comprise all the advantages of a visual dictionary model by yet including semantics in the visual words.

\subsection{Video Representation}
\label{sec:videorepresentation}
In the literature, there are many works that obtain high-level representations for videos~\cite{WACV_2016_Agharwal, CIKM_2013_Dalton, ICMR_2013_Habibian, MM_2013_Mazloom, TMM_2012_Merler, ICMR_2012_Penatti, CVPR_2012_Sadanand, CVPR_2014_Wu}.
Action Bank~\cite{CVPR_2012_Sadanand} is a method to represent videos according to a bank of action detectors. Each video is thus represented by its activations to the action bank.
Bag of Scenes~(BoS)~\cite{ICMR_2012_Penatti}, originally proposed for video geocoding, uses visual codebooks based on whole scenes instead of local patches (e.g., corners or edges). Such scenes represent places of interest, therefore, for geocoding, the BoS vector works as a place activation vector. 

Some existing methods are based on obtaining frame-level feature vectors and then aggregating them as a video vector~\cite{AAAI_2013_Jiang, CVPR_2013_Ma, MM_2013_Mazloom, CVPR_2015_Ng, MM_2015_Ye, CVPR_2015_Xu, MM_2015_Wu, BMVC_2015_Zha}.
Such methods, which also includes the BoA model, can make use of the great variety of works that aim to obtain semantic representations for images~\cite{TPAMI_2014_Bergamo, NIPS_2010_Li, IJCV_2014_Patterson}. 
For video retrieval, Mazloom~et~al.~\cite{MM_2013_Mazloom} use average pooling to aggregate frame vectors obtained by concept detectors.
Jiang~et~al.~\cite{AAAI_2013_Jiang} make use of Object Bank~\cite{NIPS_2010_Li} and Classemes~\cite{TPAMI_2014_Bergamo} to obtain the high-level frame vectors, which are then aggregated as the video vectors.
Some of such works~\cite{CVPR_2015_Xu, MM_2015_Wu, BMVC_2015_Zha} make use of CNNs for extracting frame-level features, however, they allow the use of CNN layers that do not have explicit semantics (e.g., fully-connected or even convolutional layers). Therefore, they are conceptually different from our work. For BoA, the only CNN layer that makes sense is the last layer, which is the only one with explicit semantics.

Except for BoS, all the other works mentioned above do not deal with high-level video representations in terms of high-level visual codebooks as we do in the BoA model.
Unlike BoS which uses no information from labels or from classifiers, BoA explicitly uses the labels given by supervised classifiers as feature vectors for video frames, which are then aggregated in the BoA vector.

\subsection{Video Event Retrieval}
\label{sec:eventretrieval}
A comprehensive review of event-based media processing and analysis methods can be found in~\cite{IVC_2016_Tzelepis}. 
Most of existing research works has focused on video event detection or video event recounting. 
In this work, we are interested in video event retrieval. 
Although related, there are substancial differences between such tasks~\cite{IJCV_2016_Douze}.

In~\cite{CVPR_2013_Revaud}, frame-based features are encoded into a temporal representation for videos.
Initially, dense SIFT descriptors were extracted from video frames and then PCA was used to reduce them to 32 dimensions.
Next, the reduced SIFT descriptors of each frame were encoded into a frame vector using MultiVLAD.
After that, two different approaches for encoding frame vectors into a video representation were evaluated.
The former, called Mean-MultiVLAD~(MMV), aggregates the frame vectors over the entire video using average pooling.
The latter, called Circulant Temporal Encoding~(CTE), creates a temporal representation that jointly encodes the frame vectors in the frequency domain.
In the experiments, both methods are combined, creating the MMV+CTE representation.

\begin{figure*}[!htb]
	\centering
	\iffinal
	  \includegraphics[width=\textwidth]{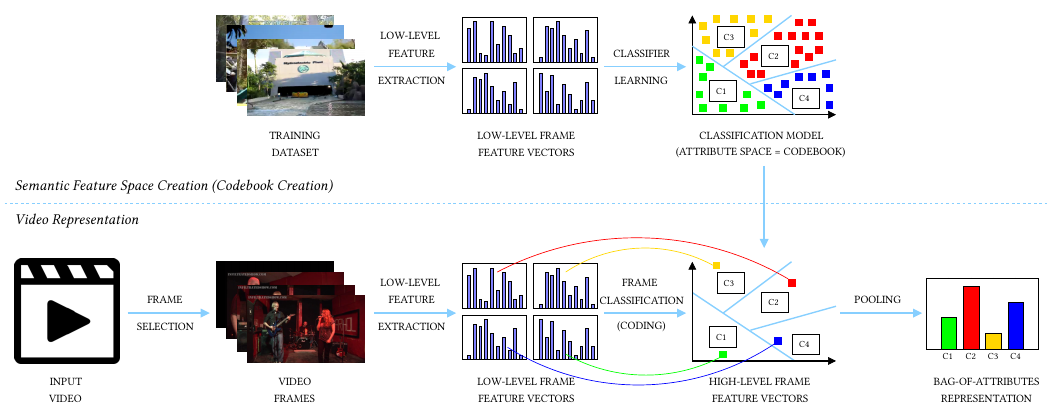}
	\else
	  \psfragfig[width=\textwidth]{pics/BagOfAttributes}
	\fi
	\caption{Overview of the Bag-of-Attributes model. On top, we show how to obtain a semantic feature space for then using it to represent videos. At the bottom, we show how to map video content into this semantic space. As the process is similar to the creation and use of visual codebooks, we also show the names that are commonly used in that scenario. $CN$ represent the categories (classes or attributes) in the training dataset.}
	\label{fig:boamodel}
\end{figure*}

In~\cite{ICCV_2013_Douze}, different hyper-pooling strategies are proposed to encode frame-level feature vectors into a video-level feature vector. 
The underlying idea relies on two stages: (i) hashing, where frame vectors are distributed into different cells; and (ii) encoding, where frame vectors in each cell are aggregated by accumulating residual vectors.
Each frame was represented using the same approach adopted in~\cite{CVPR_2013_Revaud}, as described above.
Four different hashing functions were evaluated: k-means, partial k-means (PKM), sign of stable components (SSC) e KD-Tree.
The best result was obtained with SSC.

\section{Bag of Attributes}
\label{sec:boamodel}
In this section, we present the Bag-of-Attributes (BoA) model for video representation. 
The main objective of the BoA model is to represent videos in a feature space with semantic information, resulting in a high-level representation~\cite{ICMR_2012_Penatti, NIPS_2010_Li}. 
For that, we basically need to have a semantic feature space and a mapping function from the original video space to this new space.
The steps involved in the BoA model are presented in Figure~\ref{fig:boamodel}.

In the BoA model, we obtain the semantic feature space by training a supervised classifier based on a labeled image dataset. 
The learned classifier, thus incorporates semantic information learned from the dataset. 
We call each label of the learning set as an \textit{attribute}, aiming at referring to elements containing semantics.

For mapping or coding the video properties in this semantic (high-level) feature space, we start by classifying each frame of the input video with the learned classifier.
Therefore, each frame is represented by its classification results, creating a high-level feature vector. 
Such results can be simply the class label given by the classifier or the whole probability vector, containing the probabilities of that frame in relation to every attribute of the learned classifier. 
Then, after having a high-level feature vector for each video frame, we generate the final video representation by computing some statistical measure over the frame vectors.

An obvious but important remark about the low-level feature extraction from video frames: 
in both stages (creation of the semantic space and video representation, i.e., top and bottom parts of Figure~\ref{fig:boamodel}), the low-level feature space must be the same.
For instance, if the classifier was trained with frames represented by color histograms, the classifier, of course, can only be applied over color histograms. 
Therefore, in this example, the frames of the video to be represented by BoA must have color histograms as low-level feature vectors.

We can easily map the steps in the BoA model to the steps involved in the context of visual dictionaries and bags of visual words.
The learned classifier can be seen as the \textit{codebook}: each visual word is an attribute, i.e., a region in the classification space. 
The process of classifying each frame with the learned classifier can be seen as the \textit{coding} step (visual word assignment). 
If in this step we consider only the classifier final attribution, i.e., class label for the frame, we have something similar to hard assignment. 
If we consider the classifier probability vector, we have something similar to soft assignment~\cite{TPAMI_2010_Gemert, ICCV_2011_Liu}. 
Then, the final step of summarizing the frame representations can be seen as the \textit{pooling} step, which can be implemented by summing, averaging or considering the maximum probability score among frames for each class~\cite{CVPR_2010_Boureau}.

\begin{table*}[!htb]
\centering
\caption{EVVE events list. The dataset has a total of 620 query videos and 2,375 database videos divided into 13 events. Q refers to the number of queries, Db+ and Db- are the numbers of positive and negative videos in the database, respectively.}
\label{tab:evve-event-list}
\begin{tabular}{c|l|c|c|c}
\hline
\textbf{ID}  & \textbf{Event name}                                & \textbf{Q} & \textbf{Db+}  & \textbf{Db-}  \\ \hline
1            & Austerity riots in Barcelona, 2012                 & 13               & 27            & 122           \\ 
2            & Concert of Die toten Hosen, Rock am Ring, 2012     & 32               & 64            & 143           \\ 
3            & Arrest of Dominique Strauss-Kahn                   & 9                & 19            & 60            \\ 
4            & Egyptian revolution: Tahrir Square demonstrations  & 36               & 72            & 27            \\ 
5            & Concert of Johnny Hallyday stade de France, 2012   & 87               & 174           & 227           \\ 
6            & Wedding of Prince William and Kate Middleton       & 44               & 88            & 100           \\ 
7            & Bomb attack in the main square of Marrakech, 2011  & 4                & 10            & 100           \\ 
8            & Concert of Madonna in Rome, 2012                   & 51               & 104           & 67            \\ 
9            & Presidential victory speech of Barack Obama 2008   & 14               & 29            & 56            \\ 
10           & Concert of Shakira in Kiev 2011                    & 19               & 39            & 135           \\ 
11           & Eruption of Strokkur geyser in Iceland             & 215              & 431           & 67            \\ 
12           & Major autumn flood in Thailand, 2011               & 73               & 148           & 9             \\ 
13           & Jurassic Park ride in Universal Studios theme park & 23               & 47            & 10            \\ \hline
\textbf{All} & \textbf{\textgreater\textgreater\textgreater}      & \textbf{620}     & \textbf{1252} & \textbf{1123} \\ \hline
\end{tabular}
\end{table*}

Some interesting properties of the BoA representation are: 
(i) one dimension for each semantic concept, 
(ii) compactness (dimensionality equal to the number of classes in the learned classifier), 
(iii) flexibility to use any kind of classifier for creating the semantic feature space, and 
(iv) ability to encode multiple semantic concepts in a single representation. 
The last property can be understood if we consider that in the pooling operation we keep probability scores of the multiple classes activated over the video frames. 
For instance, if our attribute space is based on objects (like the object categories of ImageNet~\cite{IJCV_2015_Russakovsky}), each frame will be classified considering the presence or not of such objects in the frame. 
The final video vector will then contain information of the objects that appear along the video. 
The BoA representation can be generalized to many other scenarios, which depend only on the attribute space to be considered. 
Other possible examples could be by considering classifiers trained to categorize scenes, faces, plants, vehicles, actions, etc.

For implementing the BoA model, different approaches can be used.
For video frame selection, techniques like sampling at fixed-time intervals or summarization methods~\cite{ISM_2010_Almeida, PRL_2012_Almeida} could be employed.
For creating the attribute classifier (i.e., the codebook), which is one of the key steps in the BoA model, one can learn the classifier in the dataset which better represents the contents of interest. 
Other option is to employ existing pre-trained classifiers, like the state-of-the-art classifiers based on CNNs~\cite{CVPR_2016_He, CVPR_2017_Huang, CACM_2017_Krizhevsky, ARXIV_2013_Sermanet, CVPR_2015_Szegedy, CVPR_2016_Szegedy, ARXIV_2014_Simonyan}, which were trained on 1000 object categories of ImageNet dataset~\cite{IJCV_2015_Russakovsky}.
In this case, the low-level feature extraction step of the BoA model is also performed by the deep neural networks, as CNNs integrate both feature extraction and classification abilities.
Therefore, video frames can be directly used as input for the CNN and its output will be the frame high-level feature vectors, considering the use of the final classification layer, which is commonly a soft-max layer.

By considering a semantic space of object categories, videos will be represented according to the existence or not of objects along their frames.
This information can help in discriminating videos from different categories.
For instance, if we have presence of object categories like weapon, knife, blast and/or fire, the video is possibly related to violent events.

\section{Experiments and Results}
\label{sec:results}
The BoA model was validated for video event retrieval.
Section~\ref{sec:dataset} describes the dataset and the protocol adopted in the experimental evaluation. 
Section~\ref{sec:parameters} discusses the impact of parameters and options for generating the BoA representation.
Finally, a comparison with the methods discussed in Section~\ref{sec:eventretrieval} is presented in Section~\ref{sec:comparison}.

\subsection{The EVVE Dataset}
\label{sec:dataset}
Experiments were conducted on the EVVE (EVent VidEo) dataset\footnote{\url{http://pascal.inrialpes.fr/data/evve/} (As of June, 2018).}: an event retrieval benchmark introduced by Revaud~et~al.~\cite{CVPR_2013_Revaud}. 
This dataset is composed of 2,995 videos (166 hours) collected from YouTube\footnote{\url{http://www.youtube.com} (As of June, 2018).}. 
Those videos are distributed among 13 event categories and are divided into a query set of 620 (20\%) videos and a reference collection of 2,375 (80\%) videos. 
Each event is treated as an independent subset containing some specific videos to be used as queries and the rest to be used as database for retrieval, as shown in Table~\ref{tab:evve-event-list}. 
It is a challenging benchmark since the events are localized in both time and space. For instance, event 1 refers to the great riots and strikes that happened in the end of March 2012 at Barcelona, Spain, however, in the database, there are a lot of videos from different strikes and riots around the world.

EVVE uses a standard retrieval protocol: a query video is submitted to the system which returns a ranked list of similar videos. 
Then, we evaluate the average precision~(AP) of each query and compute the mean average precision~(mAP) per event.
The overall performance is assessed by the average of the mAPs~(avg-mAP) obtained for all the events. 

\subsection{Impact of the Parameters}
\label{sec:parameters}
Our experiments followed the official experimental protocol created by Revaud~et~al.~\cite{CVPR_2013_Revaud}. 
Initially, each video in the dataset was represented by a BoA vector. 
With the BoA of each video, a given query video was used to retrieve the rest database videos, which were ranked according to the cosine similarity between their BoAs. 
Finally, we used the dataset official tool to evaluate the retrieval results\footnote{\url{http://pascal.inrialpes.fr/data/evve/eval_evve.py} (As of June, 2018).}.
It is important to emphasize that we analyze only the visual content, ignoring audio information and textual metadata.

\begin{figure*}[!htb]
    \centering
	\iffinal
	  \includegraphics[width=0.7\textwidth]{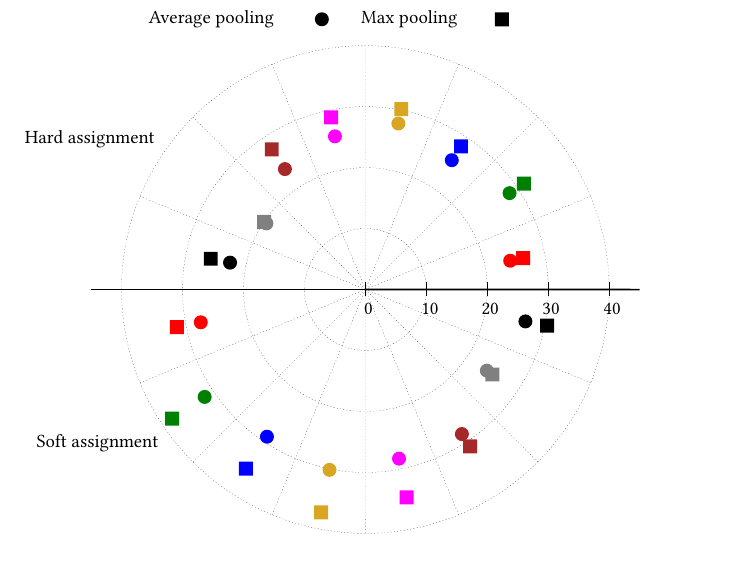} \\
	  \includegraphics[width=0.6\textwidth]{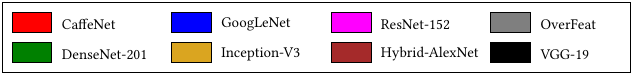}
	\else
	  \psfragfig[width=0.7\textwidth,trim={5mm 0 5mm 0},clip]{pics/parameter} \\
	  \psfragfig[width=0.6\textwidth]{pics/legend}
	\fi
    \caption{Results obtained by different configurations of BoA. The sectors refer to the CNNs used as attribute classifier. Each semicircle is related to a coding strategy. Different symbols were associated to the pooling strategies. The radius indicates the avg-mAP for each parameter setting. The farther from the center, the higher the avg-mAP. We can note that the best BoA configuration is based on DenseNet-201 with soft assignment and max pooling.}
    \label{fig:parameter}		
\end{figure*}

In this paper, video frames were selected using the well-known FFmpeg tool\footnote{\url{http://www.ffmpeg.org/} (As of June, 2018).} in a sampling rate of one frame per second. 
For obtaining the semantic feature space, we considered eight different CNN architectures: 
CaffeNet~\cite{CACM_2017_Krizhevsky}, 
DenseNet-201~\cite{CVPR_2017_Huang}, 
GoogLeNet~\cite{CVPR_2015_Szegedy}, 
Inception-V3~\cite{CVPR_2016_Szegedy},
Microsoft ResNet-152~\cite{CVPR_2016_He}, 
MIT Places Hybrid-AlexNet\footnote{\url{places.csail.mit.edu/downloadCNN.html} (As of June, 2018)}~\cite{NIPS_2014_Zhou}, 
OverFeat\footnote{\url{http://cilvr.nyu.edu/doku.php?id=software:overfeat:start} (As of June, 2018).}~\cite{ARXIV_2013_Sermanet}, and
VGG ILSVRC 19 Layers~\cite{ARXIV_2014_Simonyan}.
We used all the CNNs trained on 1000 object categories of ImageNet~\cite{IJCV_2015_Russakovsky}, except for MIT Places Hybrid-AlexNet, which considered both objects and scene categories, more specifically, 978 object categories of ImageNet and 205 scene categories of Places Database~\cite{NIPS_2014_Zhou}.
\rv{In this way, we evaluated not only the result of different classifiers, but also the effect of different types of attributes.}
Each frame was classified by such CNNs and then represented by a feature vector corresponding to the probability distribution from their softmax layer.
To obtain the BoA representation of each video, we evaluated the use of hard and soft assignment (i.e., only the predicted class or the whole probability vector, respectively) as well as average and max pooling. 

First, we conducted experiments aiming at determining the best parameter setting for the BoA model. 
Different options were evaluated for each of the steps involved in creating the BoA representation. Table~\ref{tab:parameters} summarizes the parameters and options evaluated in our experiments.

\begin{table}[!htb]
	\caption{The parameters and options evaluated for generating the BoA representation.}
	\centering
	\begin{tabular}{r|l}
		\hline
		\hline
		\textbf{Parameter} & \textbf{Options} \\
		\hline
		\multirow{3}{*}{Attribute classifier}
		& CaffeNet, DenseNet-201, GoogleLeNet, \\
		& Inception-V3, ResNet-152, Hybrid-AlexNet, \\
		& OverFeat, VGG-19 \\
		\hline
		Coding technique & hard, soft \\
		\hline
		Pooling technique & average, max \\
		\hline
		\hline
	\end{tabular}
	\label{tab:parameters}
\end{table}

Figure~\ref{fig:parameter} presents a radar plot showing the results obtained for different configurations of the BoA model. 
This chart is composed of three parts: sector, points, and radius. 
Each sector indicates the results obtained by using a given CNN as the attribute classifier.
In order to make the comparison easier, all the sectors from a same coding strategy were grouped together: the results at the top semicircle refers to the hard assignment (i.e., only the predicted class) whereas those related to the soft assigment (i.e., the whole probability vector) are presented at the bottom semicircle.
The points in each sector were associated to different symbols and each of them represents a different pooling strategy. 
Finally, the radius denotes the avg-mAP. 
The farther a point is from the center, the better the results (i.e., the higher the avg-mAP).

In general, the soft assignment (bottom semicircle) yielded better results than the hard assignment (top semicircle) for performing the coding step of BoA. 
\rv{The soft assignment takes into account the semantic ambiguity, i.e., a same frame may activate two or more attributes, which is neglected in the hard assignment, where just one attribute is activated for each frame, thus discarding relevant information about its visual content.}
With respect to the pooling strategies, max pooling (squares) yielded the highest avg-mAP scores, regardless of the attribute classifier or coding strategy used in the BoA model. 
\rv{For event retrieval, no significant difference was observed in the use of only objects (i.e., CaffeNet, DenseNet-201, GoogleLeNet, Inception-V3, ResNet-152, OverFeat, VGG-19) or both objects and scenes (i.e., Hybrid-AlexNet).}
The best results were observed using  DenseNet-201 as the attribute classifier, reaching an avg-mAP of more than 38\%.

In the next experiments, when we refer to BoA, we mean the DenseNet-201 was used for classifying video frames, soft assignment (i.e., the whole probability vector) was used for coding frame-level feature vectors, and max pooling was used to summarize them in a video-level feature vector. 
With such representation, the BoA vector contains a summary of the objects present in a video.

\subsection{Comparison with Baselines}
\label{sec:comparison}
We compared the results obtained by BoA with those reported by Revaud~et~al.~\cite{CVPR_2013_Revaud} for three baselines: Mean-MultiVLAD (MMV), CTE (Circulant Temporal Encoding) and a combination of both methods, known as MMV+CTE. 
Also, we considered the results reported by Douze~et~al.~\cite{ICCV_2013_Douze} for the variations of MMV with the following hyper-pooling functions: k-means, partial k-means~(PKM), sign of stable components~(SSC), KD-Tree and Fisher Vectors~(FV). 

\begin{figure}[!htb]
 \centering
 \iffinal
   \includegraphics[width=0.48\textwidth]{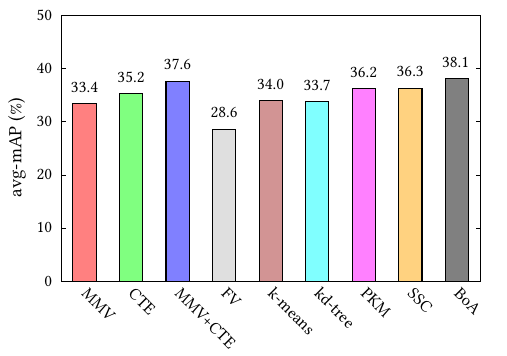} 
 \else
   \psfragfig[width=0.48\textwidth]{pics/avg-map} 
 \fi
 \caption{Performance of different methods for event retrieval on EVVE dataset. The BoA representation performed better than all the baselines.}
 \label{fig:avg-map}
\end{figure}

In Figure~\ref{fig:avg-map}, we compare the BoA representation and the baseline methods with respect to the avg-mAP. 
As we can observe, the BoA model yielded better results than all the baseline methods.

The results where also compared by event, as shown in Table~\ref{tab:map-per-event}. 
Notice that BoA performed better than the baseline methods for most of the events (7 out of 13).
For some events, the difference in favor of the BoA model is very large. 
For instance, for event 12 (\textit{``Major autumn flood in Thailand, 2011''}), the best baseline (MMV+CTE) achieves 37.1\% of avg-mAP while BoA obtains 57.1\%.

\begin{table}[!htb]
\centering
\caption{Retrieval performance (mAP) per event on EVVE dataset.}
\label{tab:map-per-event}
\begin{tabular}{c|c|c|c|c}
\hline
\textbf{Event ID} & \textbf{MMV} & \textbf{CTE}   & \textbf{MMV+CTE} & \textbf{BoA} \\ \hline

1                 & 23.9        & 13.9          & 24.6            & \textbf{32.4}        \\ 
2                 & 19.9        & 16.6          & 20.2            & \textbf{20.8}        \\ 
3                 &  8.7        & 12.8          & 11.1            & \textbf{13.8}        \\ 
4                 & 12.6        & 10.8          & \textbf{13.2}   &         12.6         \\ 
5                 & 23.4        & 26.2          & 26.0            & \textbf{29.8}        \\ 
6                 & 33.8        & 41.3          & 39.4            & \textbf{49.1}        \\ 
7                 & 12.4        & \textbf{25.2} & 21.2            &         14.9         \\ 
8                 & 25.4        & 25.7          & \textbf{28.1}   &         26.7         \\ 
9                 & 53.1        & \textbf{80.3} & 69.4            &         52.3         \\ 
10                & 45.5        & 40.9          & \textbf{48.6}   &         30.0         \\ 
11                & 77.3        & 71.4          & 77.4            & \textbf{86.4}        \\ 
12                & 36.6        & 29.7          & 37.1            & \textbf{57.1}        \\ 
13                & 60.4        & 69.3          & \textbf{71.9}   &         69.8         \\ \hline
avg-mAP           & 33.4        & 35.2          & 37.6            & \textbf{38.1}        \\ \hline
\end{tabular}
\end{table}

We also performed paired $t$-tests to verify the statistical significance of the results. 
For that, the confidence intervals for the differences between paired averages (mAP) of each category were computed to compare every pair of approaches. 
If the confidence interval includes zero, the difference is not significant at that confidence level. 
If the confidence interval does not include zero, then the sign of the difference indicates which alternative is better.

\begin{figure}[!htb]
 \centering
 \iffinal
   \includegraphics[width=0.4\textwidth]{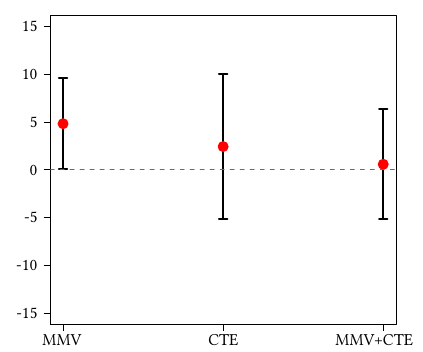} 
 \else
   \psfragfig[width=0.4\textwidth,trim={0 15mm 0 15mm},clip]{pics/paired-t-test} 
 \fi
 \caption{Paired t-test comparing BoA and the baselines. BoA outperformed MMV with statistical significance (intervals above zero), while had no significant difference to CTE and MMV+CTE (intervals cross zero), although BoA presented higher avg-mAP.}
 \label{fig:paired-t-test}
\end{figure}

Figure~\ref{fig:paired-t-test} presents the confidence intervals (for $\alpha=0.05$) of the differences between BoA and the baseline methods for the mAP measures. 
As we can observe, the confidence intervals for BoA and MMV are positive, indicating that BoA outperformed MMV. 
On the other hand, the confidence intervals for BoA and CTE include zero and, hence, the differences between such approaches are not significant at that confidence level.

We believe that BoA outperformed MMV because the BoA representation carries semantic information. 
The explicit enconding of object occurrences in the BoA representations may have created a better feature space separability for videos of different events.
On the other hand, BoA does not include temporal information and we think such feature is important to recognize some types of events. 
Unlike BoA, CTE is a more complex method that exploits the temporal information of a video.
For that reason, CTE is better than MMV and BoA for describing events composed by repeatable small sequences, like the event 9 (\textit{``Presidential victory speech of Barack Obama 2008''}).

Similar to the pooling step of the BoA model, the reasoning behind CTE is to summarize a set of frame vectors in a single video vector. 
However, different from average and max pooling, CTE is able to encode the temporal order of a frame sequence.
Therefore, in the same sense of MMV+CTE, BoA and CTE are complementary and could be combined in order to obtain a more discriminant representation.
For that, we could use CTE instead of max pooling to summarize the frame vectors in the BoA model.

The key advantage of the BoA model is its computational efficiency in terms of space occupation and similarity computation time. 
Since the time required to compute the similarity among videos is hardware dependent (i.e., with faster hardware the computational speed increases and the running time decreases) and the source codes of all the baselines methods are not available, it is challenging to perform a fair comparison of performance in relation to BoA. 
Table~\ref{tab:complexity} compares the BoA model with the baseline methods with respect to the computational complexity for similarity computation and space requirements.
In this way, we can investigate the relative difference of performance among such approaches.

\begin{table}[htb]
  \centering
  \caption{The computational complexity for similarity computation and space requirements of the different approaches in terms of the number $n$ of frames in a video$^7$.}
  \begin{tabular}{c|c|c|c}
    \hline
    \hline
    \multirow{2}{*}{\textit{Method}} & \textit{Descriptor} & \textit{Similarity} & \textit{Space} \\
	& \textit{Size} & \textit{Computation} & \textit{Requirements}\\
    \hline
	MMV & 512 & $O(1)$ & $O(1)$ \\
	CTE$^7$ & $p \times \beta \times n$ & $\Omega(n)$ & $\Omega(n)$ \\
	MMV+CTE$^7$ & $p \times \beta \times n$ & $\Omega(n)$ & $\Omega(n)$ \\
	FV & 16384 & $O(1)$ & $O(1)$ \\
	k-means & 16384 & $O(1)$ & $O(1)$ \\
	kd-tree & 16384 & $O(1)$ & $O(1)$ \\
	PKM & 16384 & $O(1)$ & $O(1)$ \\
	SSC & 16384 & $O(1)$ & $O(1)$ \\
    \textbf{BoA} & \textbf{1000} & $\boldsymbol{O(1)}$ & $\boldsymbol{O(1)}$ \\
    \hline
    \hline  
  \end{tabular}
  \\[1em]
  \parbox{0.42\textwidth}{\footnotesize{$^7$CTE relies on two parameters: the number $p$ of PQ sub-quantizers and the frame description rate $\beta$. For details regarding CTE, please, refer to~\cite{CVPR_2013_Revaud, IJCV_2016_Douze}.}}
  \label{tab:complexity}
\end{table}

As we can observe, the time complexities for both similarity computation and space occupation of CTE and MMV+CTE grows at least linear with the video length. 
For all the other methods, the computational costs are constant and, hence, they are independent of the video duration. 
Although MMV is the most compact (i.e., only 512 dimensions), its performance was one of the worst, yielding the second lowest avg-mAP of 33.4\% (see Figure~\ref{fig:avg-map}).
Clearly, BoA is much more efficient than the baselines, since it is the second most compact (i.e., less than double of MMV), but the most effective, yielding the highest avg-mAP of 38.1\%.


\section{Conclusions}
\label{sec:conclusions}
We presented a semantic video representation for video event retrieval, named Bag of Attributes (BoA). 
In this model, videos are represented in a high-level feature space, which comprises the classification space defined by a supervised image classifier.
In such space, each region corresponds to a semantic concept.
To represent video content in this space, we start by classifying each video frame with the learned classifier, resulting in a high-level feature vector for each frame (e.g., classifier probability scores).
Then, frame vectors are summarized by pooling operations to generate the final video vector, creating the BoA representation.

The main properties of the BoA representation are: each vector dimension corresponds to one semantic concept, compactness, flexibility regarding the learned classifier, and ability to encode multiple semantic concepts in a single vector.
 
To validate the BoA model for video event retrieval, we conducted experiments on the EVVE dataset.
Our implementation of BoA considered the semantic feature space created by state-of-the-art CNNs pre-trained on 1000 object categories of ImageNet. 
Such CNNs were used to classify video frames.
We evaluate the impact of different coding strategies used to encode the probability distribution from the softmax layer as high-level frame vectors.
Also, different pooling strategies were tested for summarizing the frame vectors in a final video vector (i.e., the bag of attributes). 
The results demonstrated that BoA performs similar to or better than the baselines, but using a much more compact representation. 
We believe that the ability to encode multiple concepts in the BoA representation could improve discriminating between events.

As future work, we plan to evaluate other semantic spaces created by classifiers based on CNNs (e.g., Xception~\cite{CVPR_2017_Chollet} and ResNeXt~\cite{CVPR_2017_Xie}).
In addition, we intend to evaluate different strategies for including temporal information in the representation space (e.g., using recurrent neural networks).
We also consider to evaluate classifiers trained on non-object categories, like scenes, for instance.
The evaluation of the BoA model in other applications besides event retrieval is also a possible future work.

\iffinal
\section*{Acknowledgment}
We thank the S\~{a}o Paulo Research Foundation - FAPESP (grant~2016/06441-7) 
and the Brazilian National Council for Scientific and Technological Development 
- CNPq (grants~423228/2016-1 and 313122/2017-2) for funding. We gratefully acknowledge the support of NVIDIA Corporation with the donation of the Titan Xp GPU used for this research.
\fi


\balance
%
%



\end{document}